\documentstyle[epsf]{aipproc}
\begin{document}

\title{The Massive Stellar Population in the Diffuse Ionized Gas of M33}

\author{Charles G. Hoopes \& Ren\'e A. M. Walterbos}

\address{Department of Astronomy, MSC 4500\\New Mexico State
University\\ Box 30001\\ Las Cruces, New Mexico, 88003}

\maketitle

\begin{abstract}
We compare Far-UV, H$\alpha$, and optical broadband images of the
nearby spiral galaxy M33, to investigate the massive stars associated
with the diffuse ionized gas. The H$\alpha$/FUV ratio is higher in HII
regions than in the DIG, possibly indicating that an older population
ionizes the DIG. The broad-band colors support this conclusion. The
HII region population is consistent with a young burst, while the DIG
colors resemble an older population with constant star formation. Our
results indicate that there may be enough massive field stars to
ionize the DIG, without the need for photon leakage from HII regions.
\end{abstract}


\section*{Introduction}

The majority of the ionized gas mass in galaxies exists as diffuse
ionized gas (DIG). The importance of this phase of the ISM is
emphasized by the huge amount of energy required to keep it ionized:
for example, the H$\alpha$ luminosity of the DIG in M31 is 40\% of the
total luminosity \cite{wb94}. If the DIG is photoionized, 40\% of the
ionizing photons from OB stars are required to keep the DIG ionized,
more energy than that produced by supernovae. This energy requirement
is typical of most of the galaxies that have been studied
\cite{hwg96}. Only OB stars seem to be able to provide this much
energy, but how the ionizing photons from these stars get to the DIG
is still a mystery, since DIG is seen as far as a kiloparsec away from
any HII regions in galactic disks. The question is whether the photons
from OB stars in HII regions are leaking into the diffuse medium, or
whether there are enough OB stars outside of HII regions to locally
ionize the DIG. We are addressing this problem by investigating the UV
and optical emission from the stars in HII regions and DIG in the
nearby spiral M33, and comparing to the H$\alpha$ emission from the
gas in both environments.

\section*{Data}

B, V, H$\alpha$, and red continuum images of M33 were obtained with
the 0.6 meter Burrell-Schmidt telescope at Kitt Peak National
Observatory. The 75\AA\ wide H$\alpha$ filter centered on
$\lambda$6570 also includes the [NII] $\lambda$6548 \& $\lambda$6584
lines. No correction was made for the [NII] contribution. The
H$\alpha$ image contains a total of 5 hours of integration time, and
the broadband images are 20 minutes each. The data were reduced using
standard methods. The continuum image was scaled to the H$\alpha$
image using foreground stars and subtracted. The broadband images were
calibrated using the published magnitudes for M33. The H$\alpha$ image
was calibrated using the published R magnitude and the known shape and
transmission of the continuum and line filters. The data will be
described and analyzed in Hoopes, Greenawalt, \& Walterbos
\cite{hgw97}. The Far-Ultraviolet image is a 424 second exposure taken
with the {\it Ultraviolet Imaging Telescope} \cite{s92,s97,st97},
through the B1 ($\lambda$=1520\AA, $\Delta\lambda$=354\AA) filter, and
was obtained from the public archive. All images were convolved to the
same resolution ($\sim4^{\prime\prime}$) and registered to the same
grid.

\section*{Results}

\begin{figure}[htb]
\centering \leavevmode
\epsfxsize=0.51\textwidth
\epsfbox{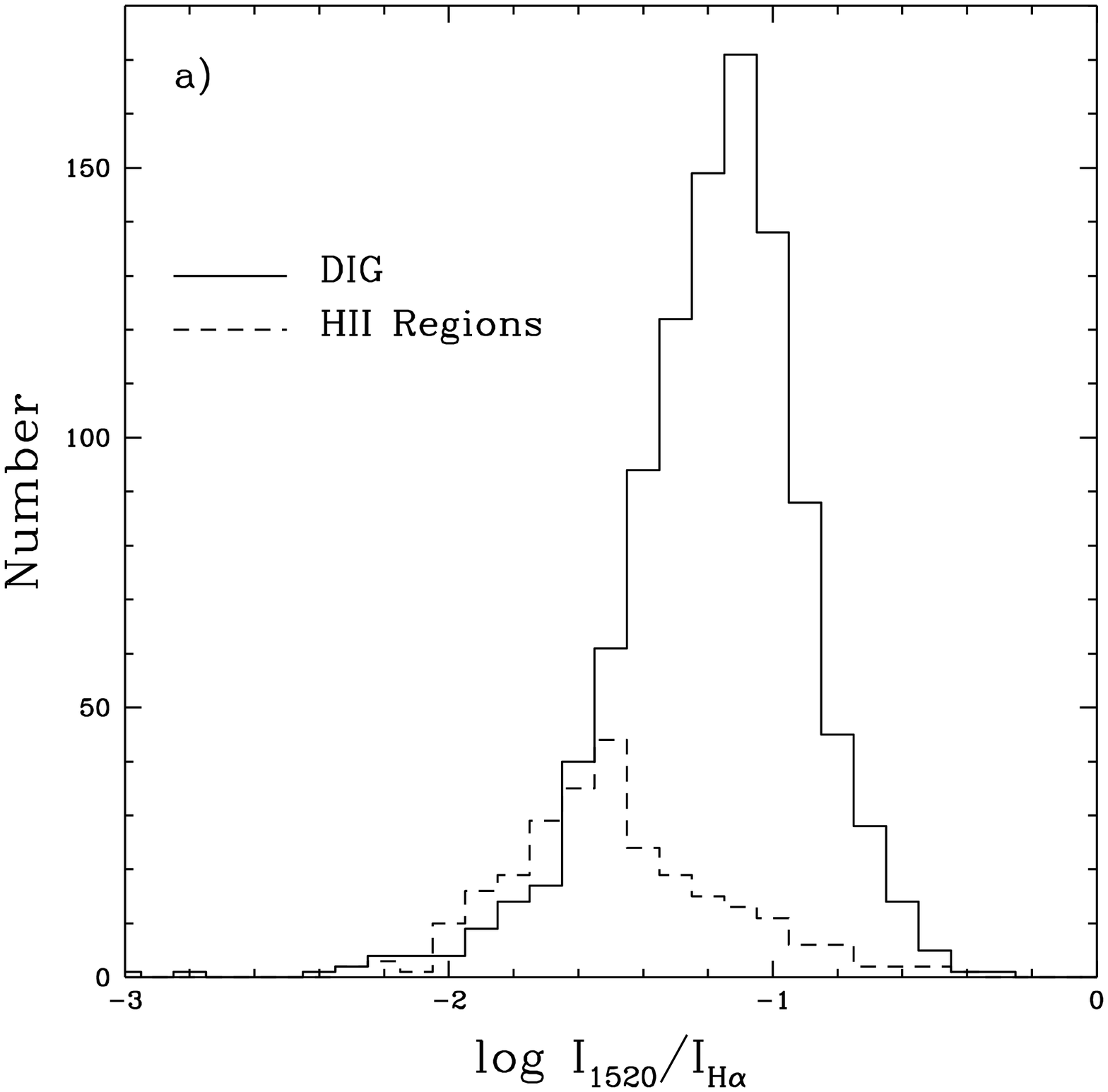}
\epsfxsize=0.51\textwidth
\epsfbox{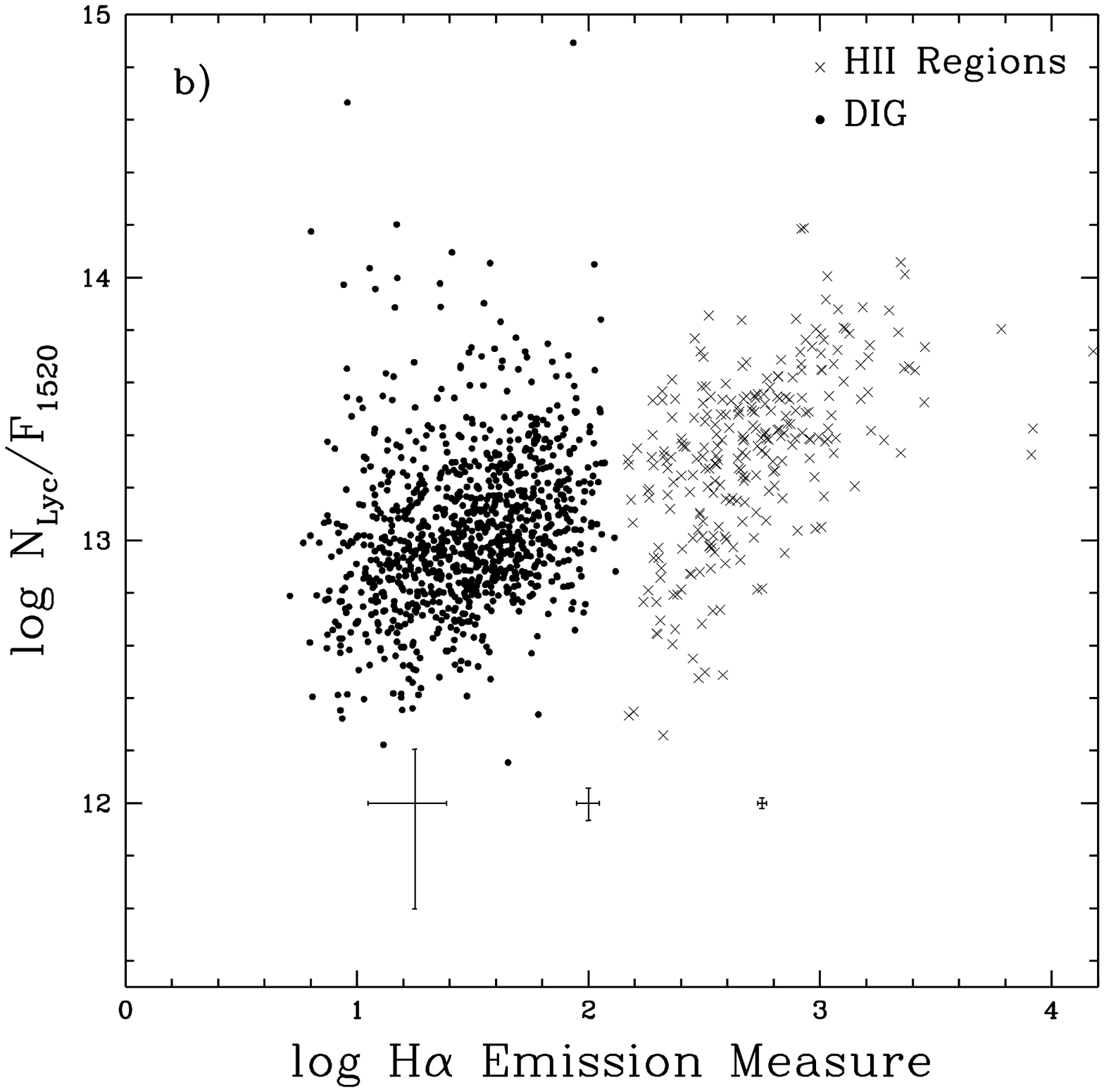}
\caption[]{ (a) Histogram of the FUV to H$\alpha$ intensities in
$20^{\prime\prime}\times20^{\prime\prime}$ (80 pc $\times$ 80 pc)
boxes centered on HII regions and DIG. The values have been corrected
for extinction using the mean HI column density in the disk of M33 and
an extinction model consisting of a uniform mixture of stars and dust,
neglecting scattering. The DIG has a systematically higher ratio,
which is consistent with later type stars dominating the radiation
field. (b) The Lyman continuum photon to FUV luminosity ratio, as a
function of H$\alpha$ surface brightness. The representative error
bars, which reflect a combination of photon noise and flat-fielding
uncertainty, are shown shifted down by a factor of 10 for clarity. The
brightest DIG has the highest ratio, which corresponds to earlier-type
ionizing stars if the ionizing photons are produced locally.}
\end{figure}

We measured the FUV and H$\alpha$ fluxes in fixed
$20^{\prime\prime}\times20^{\prime\prime}$ regions centered on DIG and
HII regions. No background was subtracted from the HII regions, so
there is a disk contribution to the flux in both bands.  The ratio of
FUV to H$\alpha$ fluxes in DIG is systematically higher than that in
HII regions (figure 1a). This indicates a difference in the average
stellar populations. If the ionizing photons are produced locally,
this could mean that the ionizing stars in the DIG are of later
spectral type than those in HII regions. Alternatively, it may mean
that the ionizing photons in the DIG are not produced locally but are
leaking from HII regions, and the FUV flux is produced by non-ionizing
B and A stars. If the difference were due to higher extinction in the
HII regions, 0.7 magnitudes A$_V$ in excess of that in the DIG would
be necessary to explain the result using a foreground screen model,
and at least 1.5 magnitudes in a uniform mixture of dust and
stars. Greenawalt, Walterbos \& Braun \cite{gwb97} found, on average,
a 0.3 magnitude difference between HII regions and DIG in M31.

The number of H$\alpha$ photons in a region is directly proportional
to the number of ionizing photons emitted by the stellar population,
assuming ionization equilibrium \cite{o89} and ignoring the effects of
dust absorption.  Figure 1b shows the ratio of Lyman continuum photons
(from the H$\alpha$ flux) to the FUV flux versus H$\alpha$
intensity. The DIG regions with the highest ratio are those with the
highest H$\alpha$ surface brightness.  Higher ratios correspond to
earlier type stars, so this may indicate that the brightest DIG, which
is usually closest to HII regions \cite{wb94}, is affected by photon
leakage, or that there is a younger field population near HII regions.

\begin{figure}[htb]
\centering \leavevmode
\epsfxsize=0.51\textwidth
\epsfbox{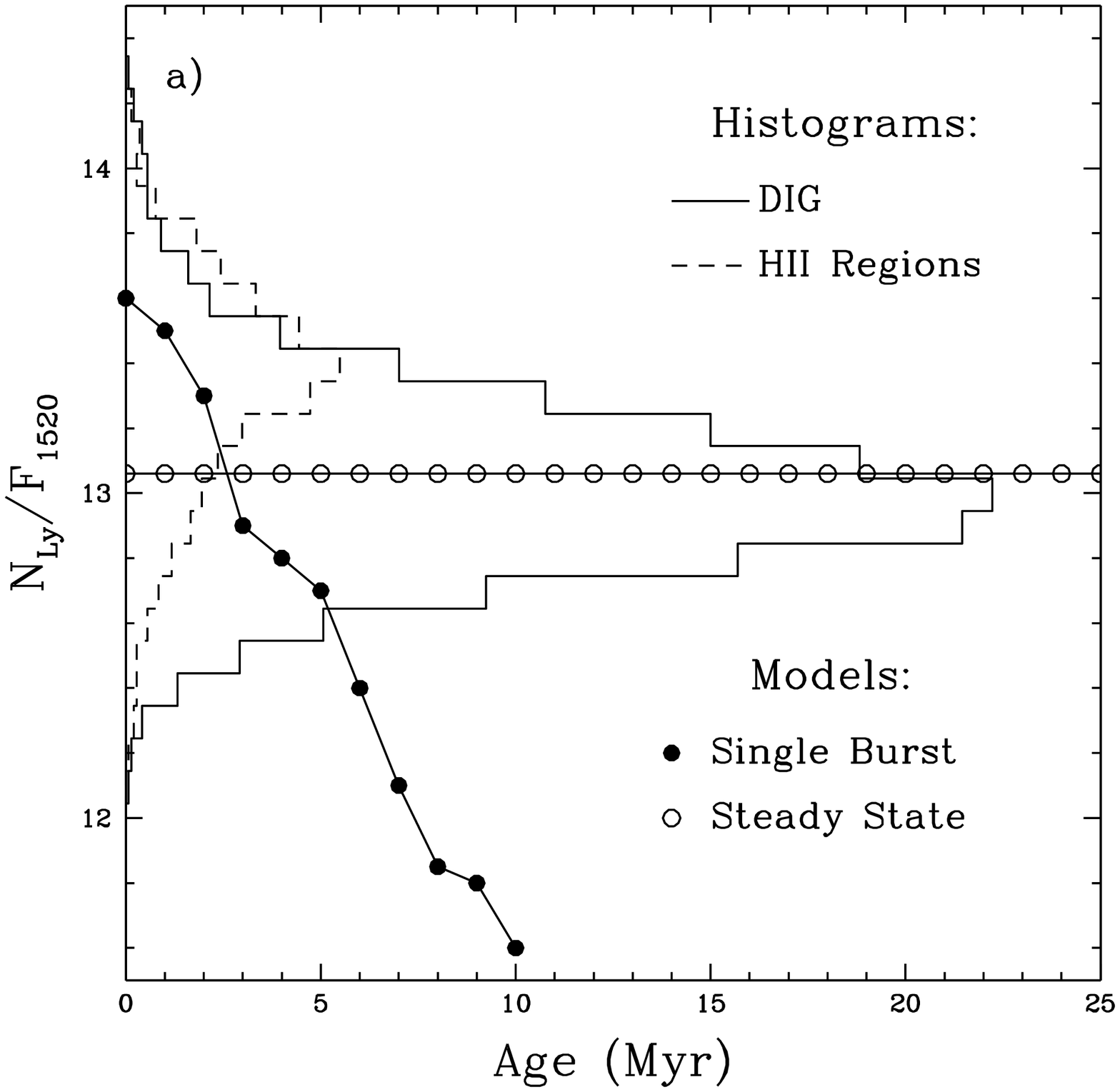}
\epsfxsize=0.51\textwidth
\epsfbox{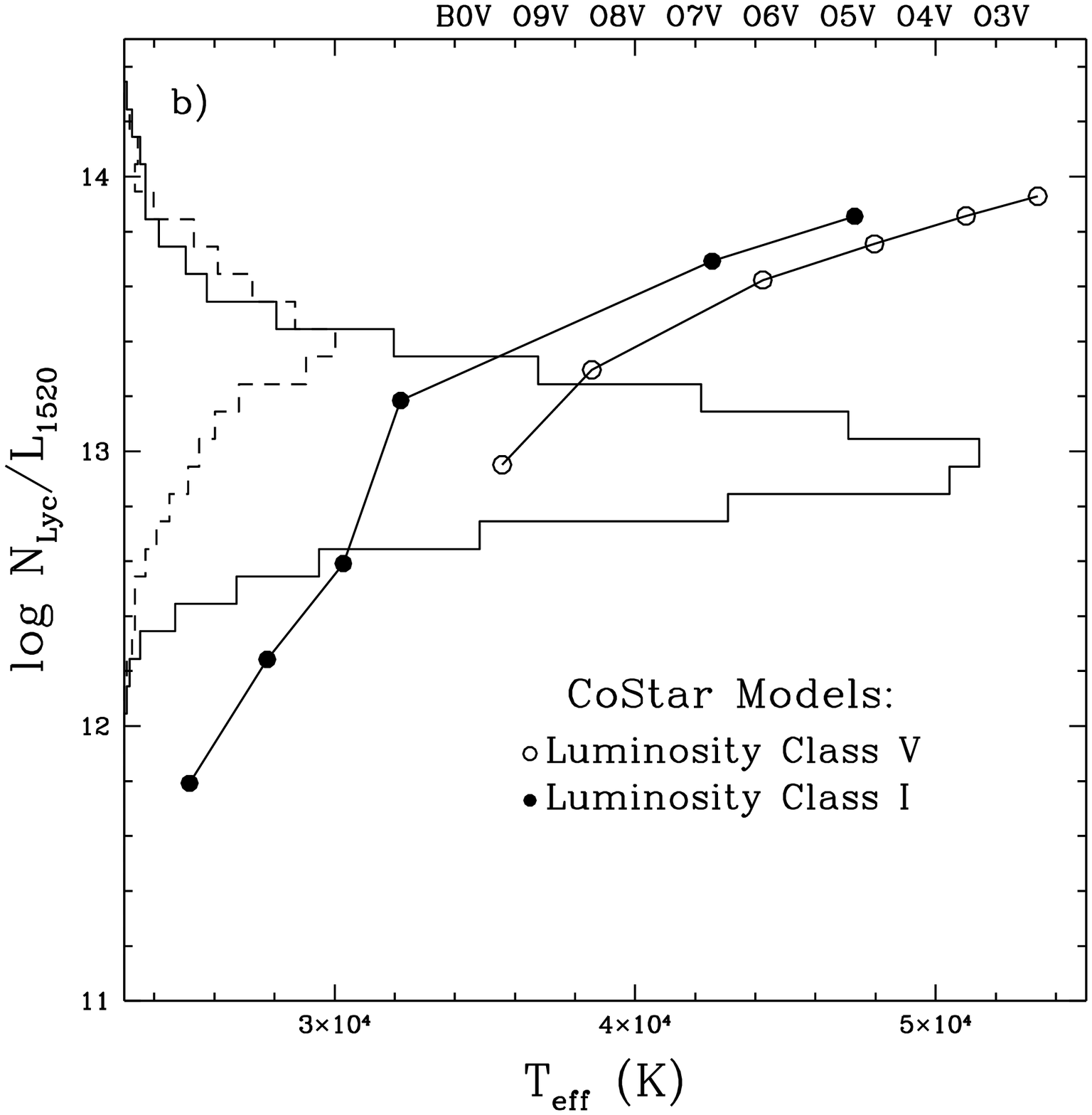}
\caption[]{ (a) Ratio of Lyman continuum photons to FUV luminosity,
compared to the evolution models presented in Hill {\it et al.}
\cite{hill95}. The histograms are the same as in figure 1a. DIG
matches an old population with constant star formation, while HII
regions are better reproduced by a young burst population, assuming
that the Lyc photons are produced locally. (b) Ratio of Lyman
continuum photons to FUV luminosity, compared to the {\it CoStar}
stellar atmosphere models of Schaerer \& de Koter \cite{sdsm96}. The
observed ratios are consistent with later-type ionizing stars in the
DIG.}
\end{figure}

\begin{figure}[htb]
\centering \leavevmode
\epsfxsize=0.51\textwidth
\epsfbox{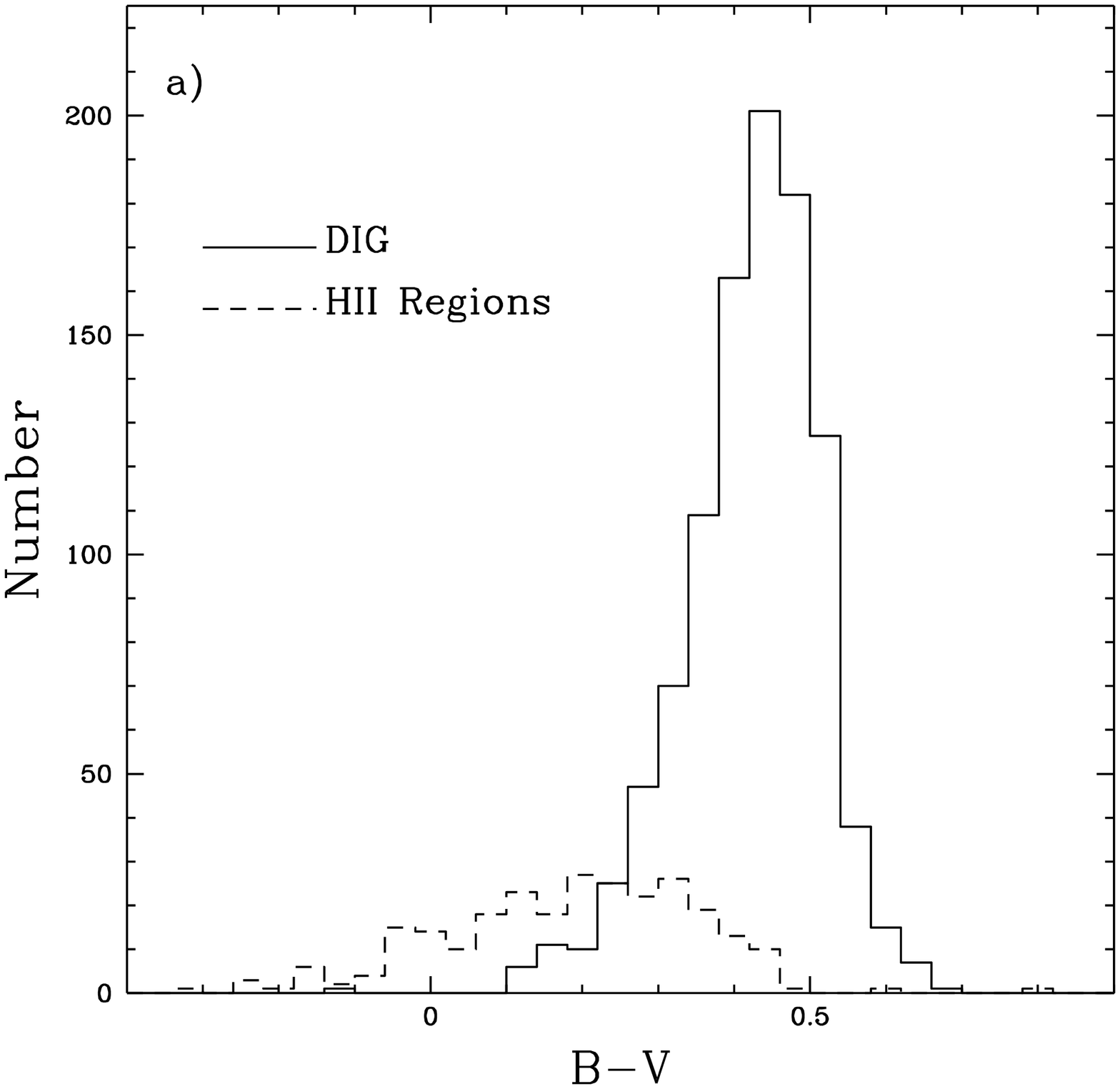}
\epsfxsize=0.51\textwidth
\epsfbox{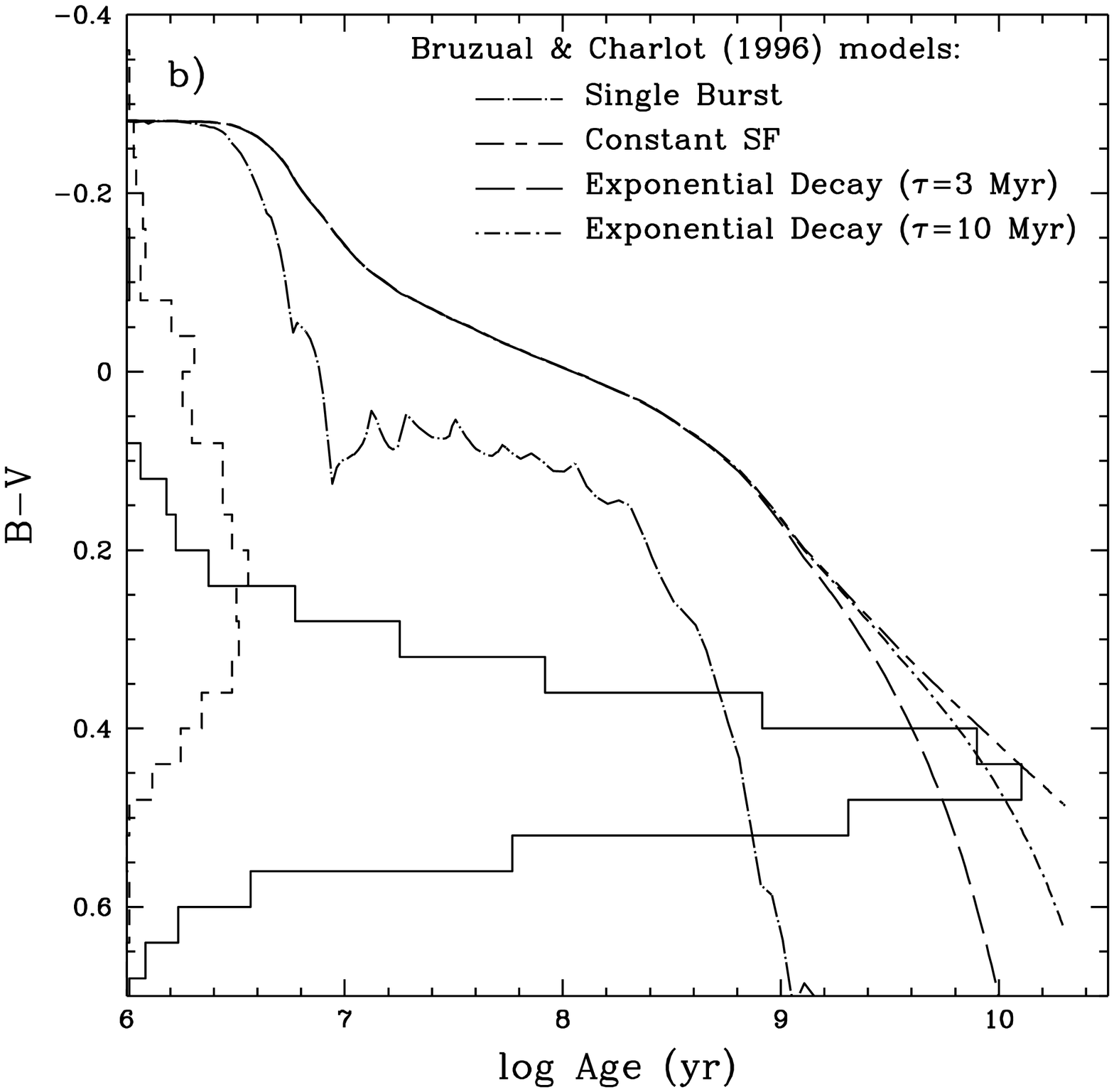}
\caption[]{ (a) Comparison of the B$-$V colors for the same DIG and
HII regions as in figure 1.  The stars in HII regions are
systematically bluer than DIG, as expected from the younger population
in the HII regions. (b) Models of evolving stellar populations from
Bruzual \& Charlot \cite{bc96}. The redder colors of the DIG are
consistent with a steady state population, with a steady inflow of new
stars. The colors of HII regions indicate a younger population. While
the DIG colors are also consistent with a $\sim$1 Gyr burst, the lack
of ionizing photons in such a population (fig 2a) rules this out. The
colors in HII regions include the contribution from the old
disk population since no background was subtracted.}
\end{figure}

\newpage  
The observed ratios can be explained with evolution models presented
in Hill {\it et al.} \cite{hill95}, as shown in figure 2a. HII regions
are described by a 0$-$5 Myr burst population, and DIG can be
described by an older burst model. However, a steady state model
reproduces the peak of the DIG distribution remarkably well. The
observed ratios in the DIG are also consistent with later-type
ionizing stars, as shown in figure 2b. If the FUV and Lyman continuum
are both produced locally, the ratio in the DIG would indicate that
B0V-O9V stars dominate the radiation field, while HII regions are
powered by O8V and earlier type stars. This is an important prediction
which can be tested by spectroscopy, using the $\lambda$5876 HeI
recombination line. Helium will be ionized by stars of type O7 or
hotter, so if this line is present in the DIG, then at least some of
the ionizing photons must come from stars in HII regions. The line was
tentatively detected in the DIG of M31 \cite{gwb97}.

The optical colors in the same 20'' $\times$ 20'' regions are
systematically bluer for HII regions, again indicating a younger
stellar population (figure 3a). Note that both regions contain a
contribution from the disk, since a local background was not
determined. This makes the colors redder than they would be for a
single population. The B$-$V colors in both environments are compared
to the evolution models of Bruzual \& Charlot \cite{bc96} in figure
3b. DIG matches an old population with a steady influx of new stars,
while the HII region colors are clearly affected by a younger
component.

\section*{Conclusions and Future Work}
 
Although the H$\alpha$-FUV comparison and FUV-optical colors of DIG
and HII regions in M33 can rule out neither Lyman photon leakage from
HII regions nor ionization by local stars, we do show that the stellar
populations in HII regions and DIG are clearly different, the colors
are consistent with a population of later-type ionizing stars in the
DIG, and the observed Lyman continuum emission relative to the FUV
flux can also be produced by these stars.  The crucial test will be to
see whether the stars that H$\alpha$-FUV analysis predicts are
actually present in the DIG. In the next step in our investigation we
will use HST FUV and optical observations to investigate the stars in
regions of DIG and HII regions in M33, and see whether the numbers and
spectral types agree with the UIT predictions. Spectroscopy of the DIG
in M33 is also planned, and will be used to search for the HeI
$\lambda$5876 recombination line. This line can test whether the
ionizing radiation field in the DIG matches the stars observed there
by UIT and HST (see figure 2b).

We thank Bruce Greenawalt for providing the H$\alpha$ image. This
research was supported by grants from the NSF (AST-9123777 and
AST-9617014), NASA(NAG5-2426), by a Cottrell Scholarship Award from
Research Corporation, and by a grant to CH from the New Mexico Space
Grant Consortium.

\end{document}